# Extended localized structures and the onset of turbulence in channel flow

J.J. Tao,[1,*] B. Eckhardt,[2] and X.M. Xiong[1]

[1]*SKLTCS&CAPT, IFSA Collaborative Innovation Center of MoE, Department of Mechanics and Engineering Science, College of Engineering, Peking University, Beijing 100871, P. R. China*

[2]*Fachbereich Physik, Philipps-University of Marburg, D-35032 Marburg, Germany*

In this letter, it is shown numerically that in plane Poiseuille flow and before the threshold of equilibrium turbulence defined by the directed-percolation universality class, a sparse turbulent state in form of localized turbulent band can sustain either by continuous increase of the turbulence fraction due to band extension when the flow domain is large enough, or by a dynamic balance between the band extension and the band breaking and decay caused by the band interaction in a finite domain. The width and tilt angle of the band keep statistically invariant during its oblique extension, a process which is not sensitive to random disturbances.

The subcritical nature of the transition to turbulence in Poiseuille flows and several other shear flows gives rise to a number of phenomena that have been documented for some time but only recently been combined in a coherent picture of the transition. Since the laminar profiles are stable against sufficiently small perturbations, turbulence can only be triggered if certain thresholds are exceeded [1-3]. Studies of optimal perturbations based on stable manifolds of critical states (so-called edge states) have shown that localized patches of downstream vortices can extract enough energy from the laminar profile and trigger turbulence [4-10]. Localized turbulent puffs in pipe flow and turbulent bands in long but narrow channels [11, 12] were found to either decay or split at low Reynolds numbers, and hence several dynamical system approaches were suggested to describe the onset of turbulence, e.g. the 1-D dynamic model [13-15], the directed percolation (DP) model [16,17], and the ecological predator-prey model [18]. Recently, the 1-D DP model was examined experimentally and numerically for Couette flows [19], where the flow is highly confined in two directions and hence the turbulent-laminar intermittency could occur only along one spatial dimension.

A natural extension of this behaviour to two spatial dimensions would suggest that localized perturbations first grow to localized spots which then grow and fill all of space. Accordingly, one would expect the transition to fall into the universality class of two spatial and one temporal dimension or (2+1)-D directed percolation. Such an expectation is consistent with recent numerical simulations of the planar shear flow between stress-free boundaries, where the statistics of the turbulent structures satisfy the power-law scaling of the (2+1)-D DP [20]. However, for planar flow with no-slip boundaries in a large domain, such as the plane-Poiseuille flow (PPF), localized turbulent bands [21] were found to survive at Reynolds numbers much lower than 830 [22, 23], the critical threshold defined by the (2+1)-D DP model [24]. Such a discrepancy is observed as well in experiments of PPF, where the turbulent fraction deviated from zero as Re<830 [24]. Exploring the underlying causes of the inconsistency between the DP model and the experimental and numerical results is the main motivation of this letter, and the answer would have consequences for the universality class of the transition.

The incompressible three-dimensional Navier-Stokes equations are solved with a spectral code [25]. Periodic boundary conditions are used in the streamwise direction $x$ and the spanwise direction $z$ and no-slip conditions are imposed at the walls ($y = \pm h$). The centerline velocity of the base state $U_c$ and the half height of the channel $h$ are chosen as the characteristic velocity and the length scale, respectively. The dimensionless flow rate is kept constant during the simulations and the Reynolds number is defined as Re=$U_c h/\nu$, where $\nu$ is the kinematic viscosity of the fluid. The spectral mode density in this paper ($n_x/L_x \geq 5.12$, $n_y/h=32$, $n_z/L_z \geq 6.4$) has been verified to be fine enough to describe the sustained turbulent flows at

Re=1000.

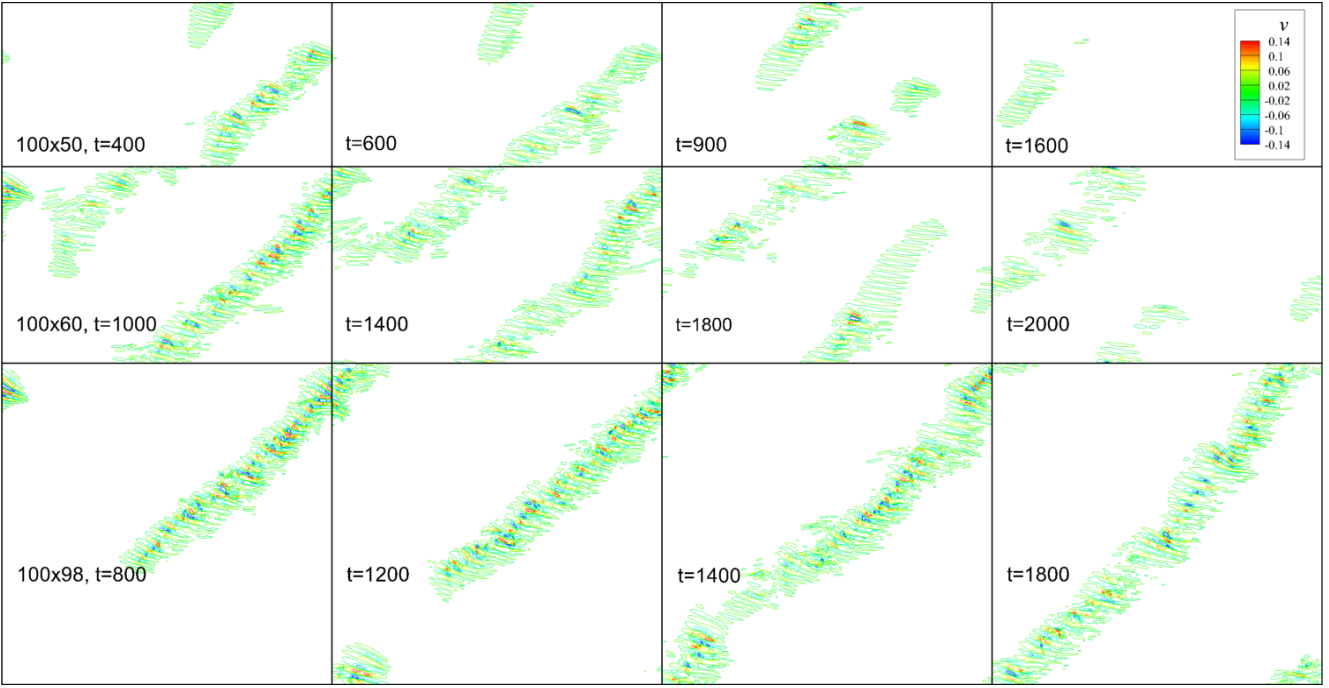

FIG. 1 Spatio-temporal evolutions of the same initial turbulent band in different domains are shown by the iso-contours of the transverse velocity $v$ at Re=700 in a moving frame with $(u,w)$=( 0.836, 0.078). The sizes of the computational domains are $L_x \times L_z$=100×50 (top), 100×60 (middle), and 100×98 (bottom), respectively.

It has been shown experimentally for PPF that the side wall or the aspect ratio of the cross section influences the onset of turbulence [26]. The conflict among the large-scale mean flows surrounding the turbulent bands may lead to band breaking and decay at low Reynolds numbers [23]. As shown in Fig.1, all calculations started with the same localized short turbulent band, and small computational domain leads to relaminarization after a transient growth by oblique extension. In the domain of $L_x \times L_z$=100×50, the band has broken by t=900 due to the self-interaction caused by the periodic boundaries and nearly vanishes at t=1600. Larger domain corresponds to weaker interaction and permits the band to grow longer, but the head and tail of the band may still interact with each other when the band is long enough (e.g. $L_x \times L_z$=100×60), leading to band breaking and decay again. For the domain of $L_x \times L_z$=100×98, whose diagonal is close to the band's statistical inclination, an interesting phenomenon occurs. The head and tail of the band connect with each other at t=1400 for Re=700 through the continuous band extension.

A natural question for the localized turbulent band is whether it will continue to extend if the computational domain is large enough. In order to describe quantitatively the characteristics of the band, we use a tilted rectangle (width $W$, length $L$, and tilt angle $\theta$) to approximate it. In the midplane, $e$ is defined as the disturbance kinetic energy relative to the base flow, and $W$, $L$, and $\theta$ can be determined as follows,

$$A_{xx} = \frac{\int ex^2 dxdz}{\int edxdz} - (\frac{\int exdxdz}{\int edxdz})^2, \quad A_{zz} = \frac{\int ez^2 dxdz}{\int edxdz} - (\frac{\int ezdxdz}{\int edxdz})^2, \quad A_{xz} = \frac{\int exzdxdz}{\int edxdz} - \frac{\int exdxdz \int ezdxdz}{(\int edxdz)^2},$$

$$\theta = \frac{1}{2}\arctan(\frac{2A_{xz}}{A_{xx} - A_{zz}}), \quad L = \sqrt{12(A_{xx} + A_{xz}\tan\theta)}, \quad W = \sqrt{12(A_{zz} - A_{xz}\tan\theta)}.$$

As shown in figure 2(a), the disturbance kinetic energy is not zero but a finite small value even $50h$ far from the band, and the large-scale mean flow provides a clockwise circulation around the band (Fig.2b), corresponding to a positive tilt angle due to a similar mechanism governing the band's orientation in plane-Couette flow [27]. The mean flow is calculated by averaging over 100 snapshots with a time step of 0.7, where the data are shifted back in the upstream direction with velocity of 0.836. The mean flow transports band-preferred perturbations to the band ends and hence contributes to the band extension. In order to describe the main body of the turbulent structure, only e>0.005 is considered in the calculation of $W$, $L$, and $\theta$. It is shown in Fig.2 (b) that the black rectangle provides a proper simplification of the band. In addition, the purple rectangle is calculated only based on $v^2$ in the midplane, lying on one side of the band and illustrating the region of small structures. Note that the purple rectangle shifts slightly to the head of the band because the transverse perturbations at the band head are statistically stronger than those in the tail region.

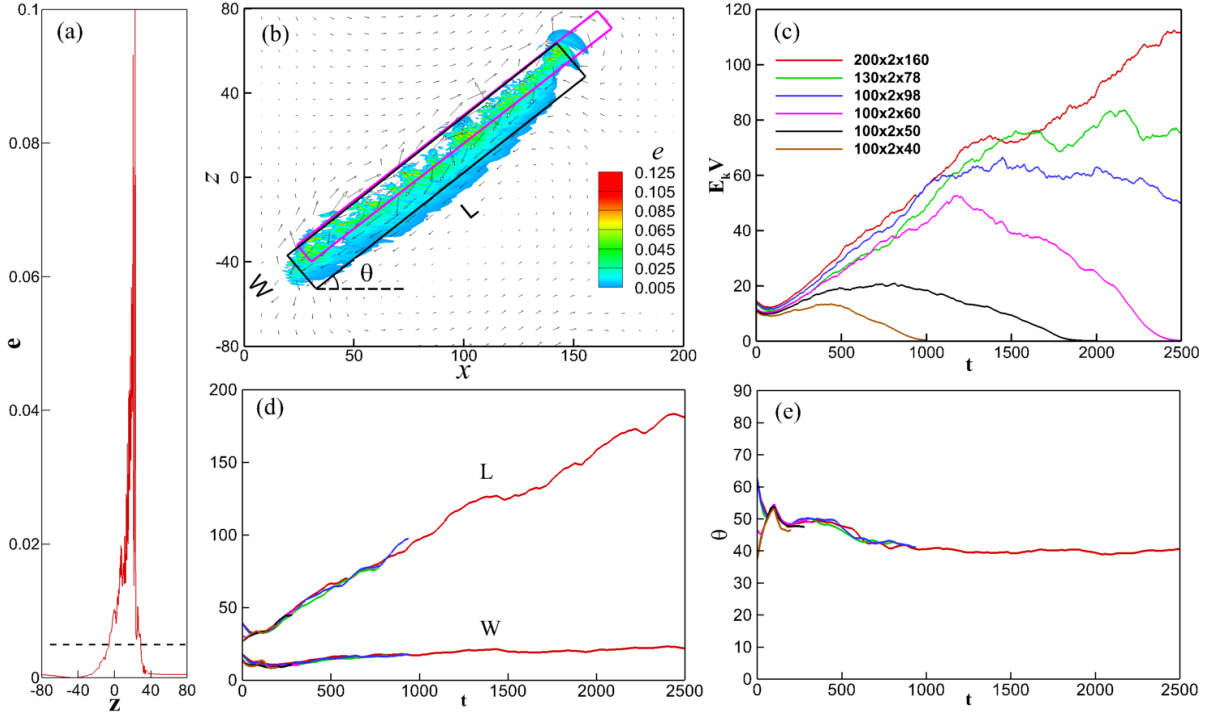

FIG. 2 Transient disturbance kinetic energy along (x,y)=(100,0) at Re=700, t=2000 is shown in (a), where e=0.005 is marked by the horizontal dashed line, and the corresponding iso-contours in the mid-plane is shown in (b). The black and the purple rectangles are calculated based on the disturbance kinetic energy in the midplane e>0.005 and $v^2$>0.0001, respectively. The vector field shows the large-scale mean flow in the plane of y=0.405 on a coarse mesh. Total disturbance kinetic energies obtained in different computational domains ($L_x \times L_y \times L_z$) at Re=700 are shown in (c), and the corresponding Length $L$, width $W$, and the tilt angle $\theta$ of the black rectangle are shown as functions of time in (d) and (e), respectively. $E_k$ is the volume-averaged disturbance kinetic energy and $V$ is the domain volume. Note that when the localized turbulent patch cannot be described by a rectangle, $L$, $W$, and $\theta$ are not calculated.

For a localized turbulent band, larger domain means less band interaction caused by the periodic boundary conditions and leads to a further increase of the total disturbance kinetic energy as shown in Fig.2(c). It is interesting to note that though the domain sizes are different, the temporal evolutions of the effective length, width, and tilt angle of the band $L$, $W$, and $\theta$ almost collapse with each other as shown in Fig.2(d) and (e), indicating that the intrinsic characteristics of the band are independent of the domain size. In addition, with the increase of $L$, $W$ and $\theta$ keep almost constant values after a short initial period, illustrating quantitatively the deterministic manner of the turbulent state. Note that the effective band width $W$ is around 20, which is much smaller than the domain size or the spacing between the band and its periodic counterpart. The total disturbance kinetic energy for domain size of $L_x \times L_y \times L_z$=100×2×60 turns to decrease around t=1200 because the head and tail of the band nearly collide with each other due to the periodic boundary conditions as shown in Fig.1. For a larger domain with the same length aspect ratio, i.e. 130×2×78, the disturbance energy continues to grow at t=1200 but turns to oscillate as t>1600 instead of decay. The reason is that the band interaction triggers band breaking, and after some broken pieces decay and the laminar regions broaden the left may grow again by oblique extension until the next band breaking occurs. Consequently, the length of the band and the corresponding kinetic energy vary with time around finite values. When the computational domain is enlarged further, e.g. $L_x \times L_y \times L_z$=200×2×160, it is shown in Fig.2 that the disturbance kinetic energy of the flow and the length of the band $L$ increase continuously following the previous trends after t=1600, suggesting that the band will continue to extend and hence can sustain if the domain is large enough. This idea is validated successfully in a very large domain of $500\times2\times400$, where the initial short band extends continuously to a single and long band of $L$=310 at t=3400 for Re=800.

It should be noted that the continuously oblique extension is not only affected by the flow-domain size but also by the Reynolds number. According to our simulations with domain size $200\times2\times160$, the volume-averaged disturbance kinetic energy $E_k$ and the effective length of the band $L$ have generally larger growth rates for higher Re, while the effective width $W$ and the band tilt angle $\theta$ are weak functions of Re. When the Reynolds number is quite low (i.e. Re<660) transient growth occurs and the localized band decays eventually. For large Re, e.g. Re>1000, turbulence spreading depends on not only band extension but also band splitting [23], and consequently the turbulence is globally sustained [28].

According to the (2+1)-D DP model, the turbulence faction will decay when Re is smaller than the critical value, which is 830 for plane-Poiseuille flow [24]. However, as illustrated above the turbulence fraction $F_t$, which may be expressed as $L\times W/(L_x\times L_z) \propto L$, can increase continuously instead of decaying at Reynolds numbers much smaller than 830 when the flow domain is large enough. For a finite domain and in a Re range below the (2+1)-D DP threshold, it is found that the localized turbulent bands can sustain as well by a dynamic balance between the band extension and the band breaking and decay caused by the band interaction. This explains why the friction factor and $F_t$ obtained in numerical simulations [23] and experiments [24] deviate from their laminar values as Re<830.

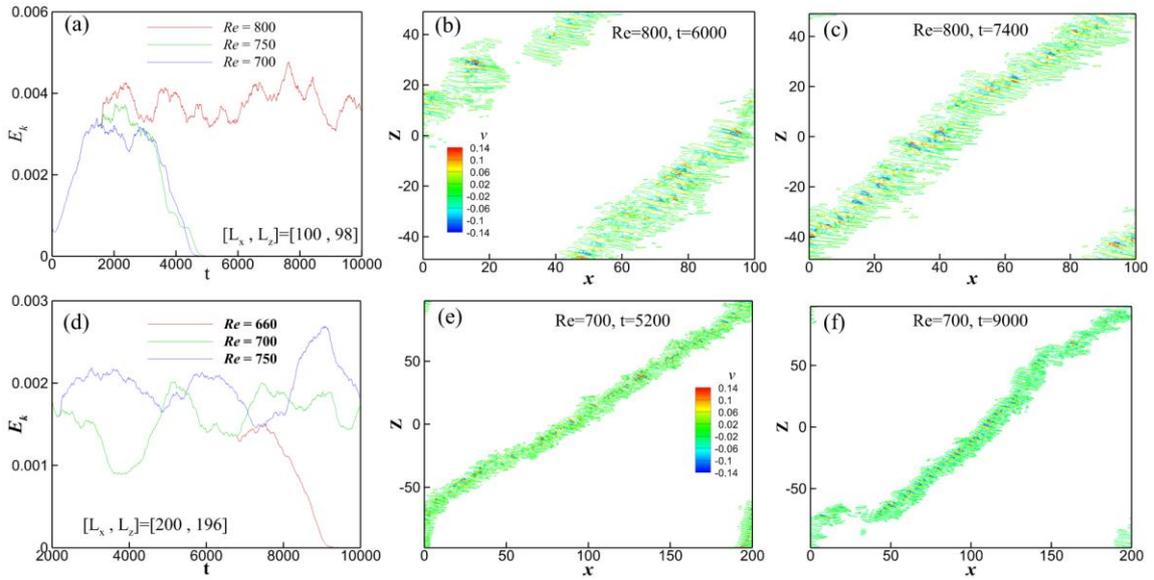

FIG. 3 Temporal series of the volume-averaged disturbance kinetic energy and the iso-contours of the transverse velocity of the periodic band obtained at different times in the midplane with $L_x\times L_z$= 100$\times$98 are shown in the upper row, where the flow of Re=700 at t=1600 is used as the initial fields for Re=750 and 800, respectively. As a comparison, the results of the periodic band in a larger domain ($L_x\times L_z$= 200$\times$196) are shown in the lower row, and the initial fields for Re=750 and 660 are the flow fields of Re=700 at t=2200 and 6800, respectively.

Based on the above discussions, a following question is whether a band suffering the band interaction can extend to be infinitely long in an infinitely large domain or how the band interaction affects the temporal evolution of the infinitely long bands. Although it is unpractical to do simulations within a domain of infinite size, the periodic band (e.g. the diagonally aligned band obtained at t=1800 for Re=700 and $L_x\times L_z$=100$\times$98 in Fig.1) reflects to some degree the properties of an infinitely long band. Assuming the x-z area scale of the flow domain ~ $l^2$, the length scale of the periodic band ~ $l$, and considering that the statistical band width $W$ is almost constant, we have $F_t \sim Wl/l^2=W/l\sim l^{-1}$. As shown in Fig.3(a), the periodic band decays at Re=700 and 750 but survives at least 10000 time units at Re=800. The volume-averaged disturbance kinetic energy for Re=800 fluctuates around a mean value about 0.0037 due to a dynamic balance between the band breaking and the band reconnection caused by the band extension as shown in Fig.3(b) and 3(c), respectively. As a comparison, a double-sized domain ($L_x \times L_z=200\times 196$) is used in the lower row of Fig.3. The solution obtained at t=1800 for $L_x \times L_z=100\times 98$ and Re=700 is extended periodically for the double-sized domain, and then the extended field is superimposed with artificial perturbations to get the flow field of the double-sized domain at t=1800. The disturbances are only added at t=1800 and are selected to guarantee that only one periodic long band will be left after a short while computation. It is shown in Fig.3(d)-(f) that when the flow-domain size is doubled and hence the band interaction is weaken, the periodic band survives at Re=700 and 750, reflecting two important characters of the initial stage during the transition. First, for a given Reynolds number, the flow domain must be large enough to sustain the periodic band, or statistically each Reynolds number corresponds to a minimal domain scale $l_{min}$ or a maximal turbulence fraction $F_{t,max}$ (note that $F_t \sim l^{-1}$). In another word, $F_t$ of the periodic band could be any value in the open interval (0, $F_{t,max}$) depending on the flow domain size, and this feature is essentially different from that of the equilibrium turbulence, whose $F_t$ has a saturated value for each Reynolds number [20]. In order to avoid confusion, the localized turbulence obtained before the equilibrium turbulence in PPF is referred as sparse turbulence hereafter. Second, larger band spacing is required for the periodic bands to survive at lower Reynolds numbers, or $F_{t,max}$ reduces with the decrease of Re. Since it is hard to find the localized band as Re<660, implying that an infinite domain is required for bands to survive and $F_{t,max}$ is infinitely close to zero, a continuous transition to the sparse turbulence is expected.

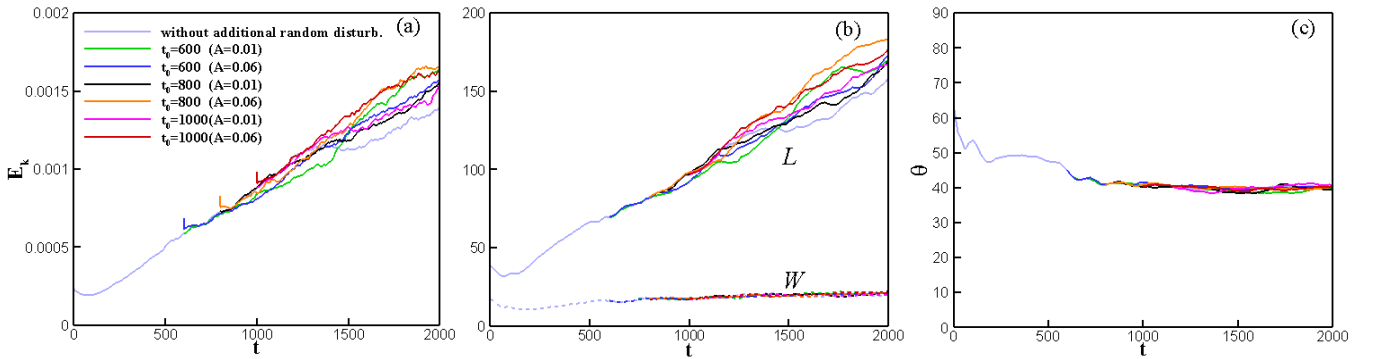

FIG. 4 Robustness of the turbulent band to random disturbances. $E_k$, $L$, $W$, and $\theta$ of the localized band as functions of time for Re=700 and ($L_x \times L_y \times L_z$)=(200×2×160) are shown in (a), (b) and (c), respectively. $t_0$ and A indicate the introduced time and the maximum amplitude of the random disturbances.

At last, we examine the robustness of the spatio-temporal process of band extension to random disturbances. As illustrated in Fig.4, the turbulent band is superimposed in the whole flow field with random disturbances of two amplitude limits (0.01 and 0.06) at three different instants (t=600, 800, 1000) respectively. The kinetic energy of all six disturbed cases increases similarly as the undisturbed one and the perturbed bands continue to extend with nearly the same characteristic width W and tilt angle θ, confirming again that the localized turbulent band own its intrinsic characteristics at the initial stage of transition.

For plane-Poiseuille flow, it is shown in this letter that before the equilibrium turbulence the localized turbulent bands form a sparse turbulence state, where the turbulence fraction doesn't decrease as expected by the (2+1)-D directed percolation model but increases continuously due to the band extension if the domain size is large enough in both *x* and *z* directions. Furthermore, different from the equilibrium turbulence where the turbulent structures will contaminate the whole field and hence the turbulence fraction will saturate at a finite value, the sparse turbulence is not space-filling and its turbulence fraction may vary with the domain size as illustrated by the periodic band. Therefore, the subcritical transition in plane Poiseuille flow, and perhaps in other spatially extended flows, follows a two-step process and is more complex than the (2+1)-D DP universality class.

The authors would like to thank many cited authors for insightful discussions. The simulations were performed on TianHe-1(A), and this work has been supported by the National Natural Science Foundation of China (Nos. 11225209, 11521091, 11490553).

*jjtao@pku.edu.cn